\documentclass[twocolumn,english,aps,prb,longbibliography]{revtex4-1}
\usepackage[T1]{fontenc}
\usepackage[latin9]{inputenc}
\usepackage{amsmath}
\usepackage{amssymb}
\usepackage{graphicx}

\makeatletter

\providecommand{\tabularnewline}{\\}

\makeatother

\usepackage{babel}
\begin{document}
\title{Strongly Correlated Transport in Topological Y-Junction Devices}
\author{E. Novais}
\date{\today{}}
\affiliation{Centro de Ci\^{e}ncias Naturais e Humanas, Federal University of ABC,
Brazil.}
\begin{abstract}
I analyze electron transport through a Y-junction formed by helical edge states
of a two-dimensional topological insulator (2DTI), focusing on the strongly
interacting regime. An experimentally motivated device geometry and a spin-conserving
tunneling Hamiltonian are proposed. I compute the conductance tensor and show that,
for specific tunneling phases and strong repulsive interactions ($g<1/2$),
transport is governed by an intermediate renormalization group fixed point
that interpolates between the weak- and strong-tunneling limits. These results
extend previous studies of point-contact tunneling and demonstrate how interactions
qualitatively modify the transport properties of multiterminal topological devices.
\end{abstract}
\maketitle

\section{introduction}

Emerging quantum technologies lie at the forefront of condensed matter
physics. Among the most promising platforms are devices that exploit
the edge states of two-dimensional topological insulators (2DTIs).
These edge states are protected by time-reversal symmetry, which suppresses
electronic backscattering. Striking examples of edge-state behavior
have been observed in Bi-based materials \citep{pauly2015subnanometrewide,yang2012spatial,drozdov2014onedimensional}
and $\text{HgTe quantum wells}$\citep{roth2009nonlocal}. Experimental
results consistently support the theoretical prediction of ballistic
transport with well-defined helicity\citep{kane2005quantum,haldane1988model}.

Technologically, it has been demonstrated that edge channels can be patterned
with nanometer precision using atomic force microscopy \citep{pauly2015subnanometrewide}
or defined electrostatically via gating techniques \citep{michetti2012tunable,michetti2013devices,mambakkam2020fabrication}.
A recent overview of potential applications of topological insulator
to nanotechnology was provided by \citet{breunig2021opportunitiesa}.
Despite of these many advances,
the practical realization of nanodevices based on these edge
states remains an open experimental challenge. 
To motivate progress in this area, it is essential that theoretical
proposals not only clarify the underlying physics but also highlight
the potential technological benefits. It is within this context that
the present study is situated.

The conventional theoretical treatment of 2DTIs typically neglects
electronic interactions. This is a reasonable first approximation,
as interaction effects are often screened and would not be noticeable
for distances larger than a scale inversely proportional to the band
gap. However, this assumption does not hold for the edge states. While
backscattering is forbidden, forward electronic scattering is symmetry-allowed
and can significantly modify edge state properties\citep{xu2006stability,hou2009corner,teo2009critical,dolcetto2016edge}.
When the edge states with opposite helicity interact, the low energy
physics is described by a Luttinger liquid. The strength and nature
of the interactions are characterized by the Luttinger parameter $g$:
repulsive interactions correspond to $g<1$, attractive interactions
to $g>1$, and noninteracting fermions to $g=1$. Experimental studies
have shown that the edge states of $\text{FeSe}$ have a Luttinger
parameter of $g\approx0.26$\citep{zhang2021tomonaga}, while monolayers
of $\text{1T}^{\prime}\text{WTe}_{2}$ span a wide range within the
strong interaction regime of $g<1/2$. Conversely, it is possible
to induce electron-electron attractive interactions ($g>1$) via the
proximity effect with a superconductor\citep{song2018spin,park2020proximity,dong2024proximityeffectinduced}.

\begin{figure}
\includegraphics[width=4cm]{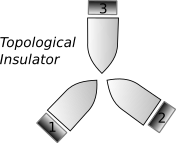}

\caption{\label{fig:Y-junction}Y-junction of three topological edge states.
The light gray regions represent three normal (non-topological) regions,
either patterned physically or induced via top gating. Edge states
(dark solid lines) with ballistic transport and well defined helicity
are at the boundary between the normal and topological regions. The
dark gray rectangles represent Fermi liquid contacts where a voltage
bias can be applied. }
\end{figure}

A significant knowledge gap remains in exploring the interplay between
topology and interactions in the transport properties of nanodevices
based on 2DTIs. An important counter example is the study of point contact
tunneling between two edge states\citep{hou2009corner,teo2009critical,dolcetto2016edge},
which has revealed intriguing results in both the weak interaction
regime ($1/2<g<2$) and in the strong-coupling regime with spin scattering\citep{teo2009critical}. 
In this work, I investigate the transport properties of interacting
edge states of topological insulators at a Y-junction\citep{yi1996quantum,nayak1999resonant,lal2002junction,chamon2003junctions,egger2003transport,oshikawa2006junctions,agarwal2009enhancement,das2006interedge,giuliano2022multiparticle}, Fig.\,\ref{fig:Y-junction}.
This is a natural
second step from the traditional point contact tunneling setup\citep{hou2009corner,teo2009critical,dolcetto2016edge}.
An experimental geometry is
proposed, Fig.\,\ref{fig:Y-junction}, and the corresponding conductance
tensor is evaluated. 

The analysis rests on several key theoretical assumptions. First,
the edge states forming the Y-junction are assumed to be well described
by their low energy field theoretical limit. A specific microscopic
geometry is considered, illustrated in Fig.\,\ref{fig:Microscopic-model-Y},
where the relative orientation of the sublattices plays a crucial
role in setting the tunneling phases; alternative geometries would
generally require modifying the phase structure accordingly. The model
also assumes that the tunneling processes conserve the spin projection
of the fermions, as is expected in $\text{HgTe}$ samples\citep{hou2009corner}.
For simplicity and symmetry, the tunneling amplitudes between any
pair of edges are taken to be identical. Additionally, the amplitudes
for correlated hopping are assumed to be positive, a reasonable assumption
if they emerge solely from second order perturbation theory. Finally,
each edge is coupled to an external Fermi liquid contact far from
the junction, as depicted in Fig.\,\ref{fig:Y-junction}.

In contrast to the usual point contact tunneling between edge states,
the most interesting physics for a Y-junction
lies in the experimentally relevant \emph{strong interaction regime}
($g<1/2$) \emph{without spin scattering}, where the conductance is dominated by
correlated hopping. The central result of this work is the identification and analysis of an intermediate
renormalization group fixed point that control the conductance behavior
in a broad range of parameters.

The manuscript is organized as follows. Section\,\ref{sec:the-y-juction}
reviews the Kane-Mele model\citep{kane2005quantum}, which serves
as a prototype for more general two-dimensional topological insulators.
This section also discusses the mean-field proximity effect to a superconductor
and presents a microscopic description of the Y-junction. Section\,\ref{sec:Luttinger-liquid-descriptions} introduces the framework
for describing the edge states and the Y-junction in terms of Luttinger
liquids. To build intuition, the non-interacting case is analyzed
in Section\,\ref{sec:Conductance-tensor-for-g1}. The weak-coupling
renormalization group equations are derived in Section\,\ref{sec:stability-of-the},
and the scattering matrix at refermionized points of the phase diagram
is obtained in Section\,\ref{sec:refermionization-at-2}. The strong-coupling
fixed points are discussed in Section\,\ref{sec:strong-tunneling-fixed},
and the conductance tensor is evaluated in Section\,\ref{sec:Y-junction-conductance}.
Finally, the manuscript concludes with a discussion and summary of
the results in Section\,\ref{sec:Discussion-and-Conclusions}.

\section{the y-juction\label{sec:the-y-juction}}

Although it is valid to begin the discussion directly with the Luttinger
liquid description of the edge states, it is pedagogically valuable
to first consider the archetypal microscopic models proposed by Haldane\citep{haldane1988model}
and Kane-Mele\citep{kane2005quantum} for 2DTI. In the manuscript
I use natural units $k_{B}$, $c$ and $\hbar=1$, but whenever necessary SI units
can be reintroduced by dimensional analyses. 

\begin{figure}
\includegraphics[width=3.5cm]{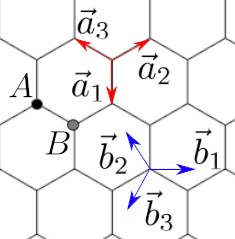}\caption{\label{fig:The-sublattices}The sublattices $A$ and $B$ of the honeycomb
lattice.}
\end{figure}

The Haldane model describes a honeycomb lattice, Fig.\,\ref{fig:The-sublattices}
, with real nearest-neighbor hopping, $t_{1}$,and complex, chiral
second-nearest-neighbor hopping, $t_{2}=\left|t_{2}\right|e^{i\varphi}$.
This setup realizes integer quantum Hall physics \emph{without} the
need for an external magnetic field. 

Kane and Mele\textquoteright s key insight was that such complex hoppings,
$t_{2}$, naturally emerges from the spin-orbit coupling in some materials.
As a result, their model effectively consists of two time-reversed
copies of the Haldane model-one for each spin projection-\emph{with
conjugate} $t_{2}$ \emph{parameters}. Consequently, the Kane-Mele
model preserves time-reversal symmetry and features edge states with
well-defined helicity in any finite sized sample.

When translational invariance holds, the model admits a particularly
simple form in momentum, $k$, and spin, $\sigma$, space
\begin{widetext}
\begin{equation}
H_{0}=\sum_{k\sigma}\left[\begin{array}{cc}
c_{A,\vec{k},\sigma}^{\dagger} & c_{B,\vec{k},\sigma}^{\dagger}\end{array}\right]\left[\begin{array}{cc}
h_{\vec{k},\sigma}+f_{\vec{k},\sigma} & g_{\vec{k},\sigma}\\
g_{\vec{k},\sigma}^{*} & h_{\vec{k},\sigma}-f_{\vec{k},\sigma}
\end{array}\right]\left[\begin{array}{c}
c_{A,\vec{k},\sigma}\\
c_{B,\vec{k},\sigma}
\end{array}\right],\label{eq:KM-model}
\end{equation}
\end{widetext}
where the nearest neighbors lattice space is set to unity, $A/B$
label the two sublattices of the honeycomb lattice, $\left\{ c_{\alpha,\vec{k},\sigma},c_{\beta,\vec{q},\sigma^{\prime}}\right\} =\delta_{\alpha,\beta}\delta_{\vec{k},\vec{q}}\delta_{\sigma,\sigma^{\prime}}$
are the standard second quantization fermionic operators, $\varphi=\varphi_{\uparrow}=-\varphi_{\downarrow}$, $h_{\vec{k}\sigma}=2\left|t_{2}\right|\cos\varphi_{\sigma}\left[\sum_{j=1}^{3}\cos\sqrt{3}\vec{k}.\vec{b}_{j}\right]$, $f_{\vec{k}\sigma}=-2\left|t_{2}\right|\sin\varphi_{\sigma}\left[\sum_{j=1}^{3}\sin\sqrt{3}\vec{k}.\vec{b}_{j}\right]$
, $g_{\vec{k}}=t_{1}\left[\sum_{j=1}^{3}e^{i\vec{k}\vec{a}_{j}}\right]$
and $\vec{a}_{j}$ and $\vec{b}_{j}$ are define on Table\,\ref{tab:lattice-def}.

The spectrum of $H_{0}$ has an energy gap 
\begin{equation}
\delta E_{\vec{k},\sigma}=2 \sqrt{f_{\vec{k},\sigma}^{2}+\left|g_{\vec{k},\sigma}\right|^{2}},\label{eq:gapH0}
\end{equation}
that has its smallest value close to two non-equivalent points at
the edge of the Brillouin zone, $\vec{k}_{D}=\left(\frac{2\pi}{3},\pm\frac{2\pi}{3\sqrt{3}}\right)$\citep{haldane1988model}.
Close to these points the gap is approximately given by $\delta E_{D}\propto\sqrt{3}\sin\varphi\left|t_{2}\right|$.
In real experimental settings, the value of the topological gap can
depend on the substrate. For instance, in bismuth bilayers can reach
up to $0.3\text{ eV}$\citep{drozdov2014onedimensional} under certain
experimental conditions.

Another important experimental control parameter is a position-dependent
chemical potential, 
\begin{equation}
H_{\gamma}=\sum_{\alpha=\left\{ A,B\right\} ,\vec{r},\sigma}\gamma\left(\vec{r}\right)c_{\alpha,\vec{r},\sigma}^{\dagger}c_{\alpha,\vec{r},\sigma},\label{eq:Hmu}
\end{equation}
which can potentially alter the topological character of the sample\citep{michetti2012tunable,michetti2013devices}.
If the local chemical potential is much larger than the topological
gap, the corresponding energy level shifts outside the gap, effective
creating a topological ``hole'' in the 2DTI. An equivalent experimental
avenue was explored in the heterostructure $InAs/GaSb$\citep{suzuki2015gatecontrolled},
where a semi-metal/topological transition was induced by a local gate.

Therefore, the topology of the sample can be engineered through geometric
modifications, such as patterning with a scanning tunneling microscope\citep{fendley2009boundary},
or by applying a voltage to a top gate. Both approaches create a region
of \textquotedblleft normal\textquotedblright{} (non-topological)
material that has helical edge states along its boundary.

In two-dimensional systems, Coulomb screening is limited, making it
natural to expect repulsive interactions among electrons in a 2DTI.
Conversely, attractive interactions can be experimentally introduced
via the proximity effect with a superconductor. The resulting changes
in the band structure can be estimated by adding a mean-field pairing
term to Eq.\,(\ref{eq:KM-model}),

\begin{equation}
H_{\Delta}=\Delta\sum_{\alpha=\left\{ A,B\right\} ,\vec{k}}c_{\alpha,\vec{k},\uparrow}^{\dagger}c_{\alpha,\vec{k},\downarrow}^{\dagger}+\text{h.c.}.\label{eq:H_delta}
\end{equation}
The total Hamiltonian $H=H_{0}+H_{\Delta}$ can be straightforwardly
solved using the Nambu formalism to write the Bogoliubov-de Gennes
equations. The resulting band structure consists of four energy bands,
with the dispersion relation for $\phi=\pi/2$ given by
\begin{equation}
E_{k}=\pm\sqrt{\Delta^{2}+f_{\vec{k}}^{2}+\left|g_{k}\right|^{2}\pm2\Delta\left|\text{\ensuremath{\Im}}\left(g_{\vec{k}}\right)\right|},
\end{equation}
where $\text{\ensuremath{\Im}}$ denotes the imaginary part of $g_{\vec{k}}$.
As long as the energy gap does not close, the system remains in the
topological phase. A condition that is satisfied when $\Delta<\left|t_{2}\right|<t_{1}$.

\begin{table}
\[
\begin{array}{|l|l|}
\hline \vec{a}_{1}=\left(0,-1\right) & \vec{b}_{1}=\left(1,0\right)\\
\vec{a}_{2}=\left(\frac{\sqrt{3}}{2},\frac{1}{2}\right) & \vec{b}_{2}=\left(-\frac{1}{2},\frac{\sqrt{3}}{2}\right)\\
\vec{a}_{3}=\left(-\frac{\sqrt{3}}{2},\frac{1}{2}\right) & \vec{b}_{3}=\left(-\frac{1}{2},-\frac{\sqrt{3}}{2}\right)
\\\hline \end{array}
\]

\caption{\label{tab:lattice-def}Definitions used for the honeycomb and triangular
lattice vectors.}
\end{table}

\begin{figure}
\includegraphics[trim={0px 5px 0px 18px},clip,width=1.05\columnwidth]{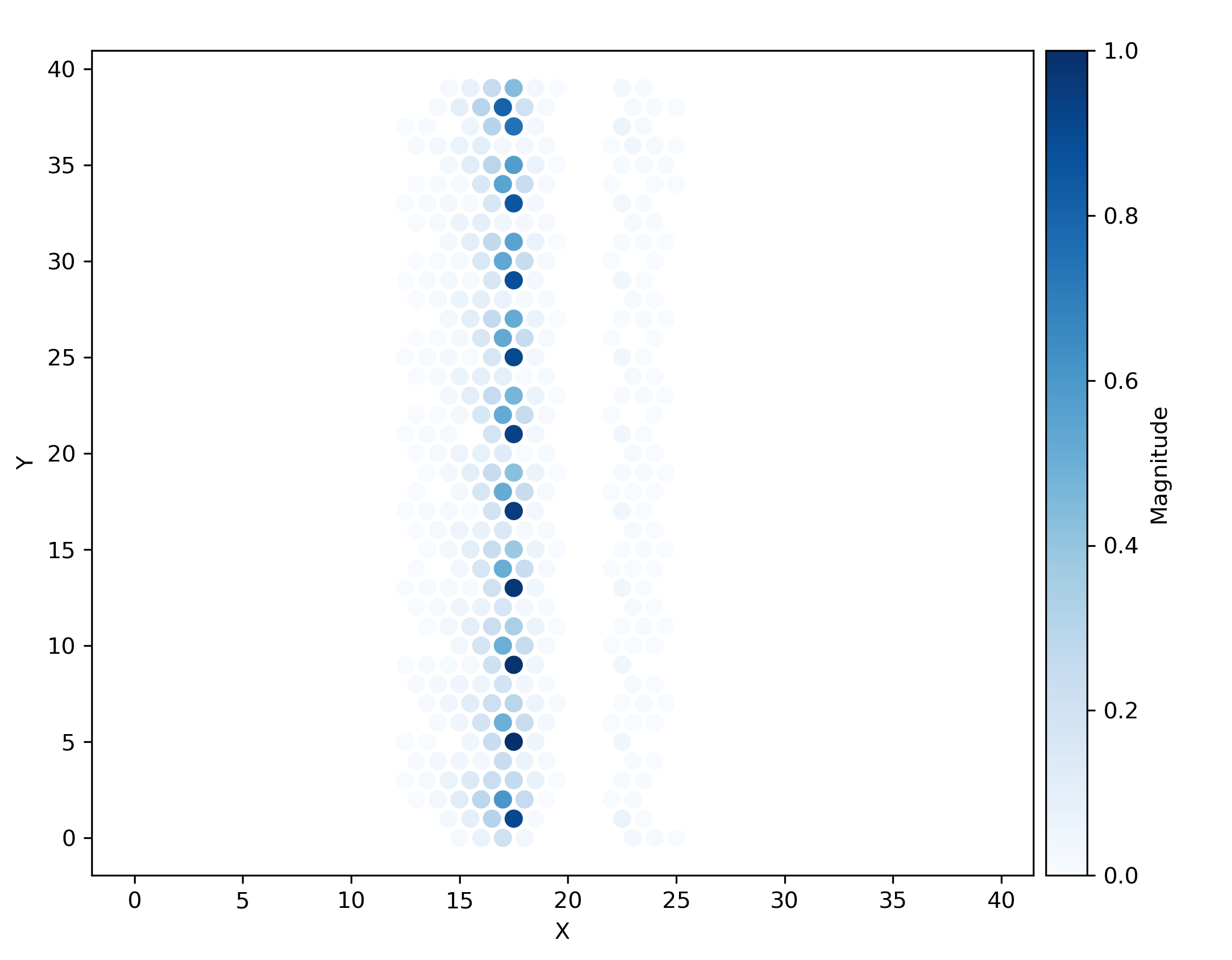}

\caption{\label{fig:Tight-biding-electronic-density}Tight-biding electronic
density for an edge state by exact diagonalization. The simulated
lattice contains 40 unit cells in each spatial direction, with parameters:
$\left|t_{2}\right|/t_{1}=0.5$, $\Delta/t_{1}=0.2$, and $\varphi=\pi/2$.
The five central columns of sites have a chemical potential
$\gamma\left(\left|\vec{r}\right|\right)/t_{1}=3$, that moves the tight-biding
states out of the electronic band. The plot shows the electronic density
 on sublattice $B$ and spin projection $\downarrow$.}

\end{figure}

All these elements are illustrated in Fig.\,\ref{fig:Tight-biding-electronic-density}.
The figure shows the wave-function probability density of an edge state,
$\left|\psi_{\sigma}\left(\vec{r}\right)\right|^{2}$, obtained by
exact diagonalization of the full Hamiltonian $H=H_{0}+H_{\gamma}+H_{\Delta}$
on a finite lattice with periodic boundary conditions. The simulated
system contains 40 unit cells in each spatial direction, with parameters:
$\left|t_{2}\right|/t_{1}=0.5$, $\Delta/t_{1}=0.2$, $\varphi=\pi/2$,
and a spatially varying chemical potential $\gamma\left(\left|\vec{r}\right|\right)/t_{1}=3$
for all sites in the five middle columns and zero otherwise.

The proposed device is defined in Fig.\,\ref{fig:Y-junction}. It
consists of three normal (non-topological) regions, either physically
patterned or induced via top gating. Each region is bounded by a pair
of counter-propagating edge states with opposite helicities. Far from
the central region, three metallic contacts are modeled as normal
Fermi liquids, from which electrons can tunnel into the nearest edge.

\begin{figure}
\includegraphics[width=5cm]{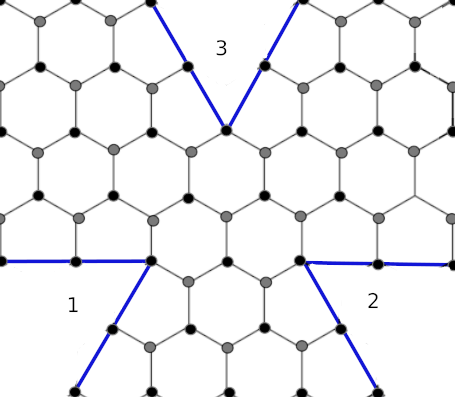}

\caption{\label{fig:Microscopic-model-Y}Microscopic
description of the $Y$-junction on a topological insulator using
the Kane-Mele model. The blue line represent edge states with define
helicity, for instance: up spins propagating clockwise and down spins
propagating counterclockwise.}
\end{figure}

Fig.\,\ref{fig:Microscopic-model-Y} presents a close-up of the region
where the three edge states are the closest to each other-referred
to here as the ``Y-junction''. To model the junction, I assume that
each edge state lies on the same sublattice and forms a corner at
the point nearest to the junction center. Although these assumptions
are not essential, they simplify the calculations and allow for a
clear identification of the tunneling phases between different edge
states. Finally, since a topological insulator respects time-reversal
symmetry, the $Y$ junction made by its edge states must also preserve
this symmetry.

\section{Luttinger liquid descriptions of the edge states\label{sec:Luttinger-liquid-descriptions}}

The edge-state wave function decays exponentially in the direction
perpendicular to the edge. The penetration depth of a given state
depends on its energy and is smallest for states near the center of
the topological gap. To a very good approximation, the low-energy
dynamics of electrons along each edge can be described by Dirac fermions
with well defined helicity. Accordingly, the Hamiltonian for the $j$-th
edge is given by

\begin{equation}
H_{\text{free edge}}^{\left(j\right)}=iv_{0}\int_{-\infty}^{\infty}dx\psi_{\uparrow,j}^{\dagger}\partial_{x}\psi_{\uparrow,j}-\psi_{\downarrow,j}^{\dagger}\partial_{x}\psi_{\downarrow,j},\label{eq:fermionic_edge}
\end{equation}
where $\uparrow/\downarrow$ denote left/right moving fermions (and
also spin projection) and $v_{0}$ is the fermionic velocity. The
labeling of right/left movers is arbitrary, since it corresponds to
a coordinate choice. Eq.(\ref{eq:fermionic_edge}) assumes that the
normal regions defining the ``Y'' junction are sufficiently large
to allow the field theory description of the edge states. The one-dimensional
parameterization of the edge is chosen so that the origin, $x=0$,
corresponds to the corner point\citep{hou2009corner}. This convention
follows previous studies on edge state junctions\citep{hou2009corner,teo2009critical,dolcetto2016edge}.

Since the edge states have an exponentially small width, the electronic
transport between different edges is dominated by tunneling between
their closest points. The single electron tunneling operator between
edges $j$ and $k$ is therefore given by

\begin{align}
H_{e}^{\left(j,k\right)}= & \lambda_{e}\left[e^{\pm ir\varphi}\psi_{\uparrow,j}^{\dagger}\left(0\right)\psi_{\uparrow,k}\left(0\right)\right.\nonumber \\
 & \left.+e^{\mp ir\varphi}\psi_{\downarrow,j}^{\dagger}\left(0\right)\psi_{\downarrow,k}\left(0\right)\right]+\text{h.c.},\label{eq:eq:fermionic-tunneling}
\end{align}
where $r=  d_0  / \sqrt{3}a$, $d_0$ is the distance between the edges, and $\varphi$
is the same phase from the Kane-Mele model (on Fig.\,\ref{fig:Microscopic-model-Y}
$r=2$). I assume that tunneling preserves the spin projection, as
expected in $\text{HgTe}$ samples\citep{hou2009corner}. For simplicity,
all tunneling amplitudes, $\lambda_{e}$, are identical. 

As discussed in \citet{hou2009corner}, the ``$g_{2}$'' forward
scattering is the only interaction that is consequential to the dynamics
of the edge states. This interaction corresponds to a density-density
interaction between fermions of opposing helicity,

\begin{equation}
H_{h_{2}}^{\left(j\right)}=g_{2}\int_{-\infty}^{\infty}dx\psi_{\downarrow,j}^{\dagger}\psi_{\downarrow,j}\psi_{\uparrow,j}^{\dagger}\psi_{\uparrow,j}.\label{eq:fermionic-interaction}
\end{equation}

The total Hamiltonian for the $j$-th edge is 
\begin{equation}
H_{\text{edge}}^{\left(j\right)}=H_{\text{free edge}}^{\left(j\right)}+H_{h_{2}}^{\left(j\right)},
\end{equation}
and can be conveniently analyzed using standard Abelian bosonization techniques\citep{delft1998bosonization,gogolin2004bosonization}.
The bosonic identities are usually written as
\begin{align}
\psi_{\uparrow j} & =\eta_{\uparrow j}\frac{e^{-i\phi_{\uparrow j}}}{\sqrt{\varepsilon}},\label{eq:bosonic-identity-up}\\
\psi_{\downarrow j} & =\eta_{\downarrow j}\frac{e^{i\phi_{\downarrow j}}}{\sqrt{\varepsilon}},\label{eq:bosonic-identity-down}
\end{align}
where $\varepsilon$ is a short time cut-off and $\eta_{\alpha}$
are the Klein factors that ensuring anticommutation relations between
different fermionic species.

Klein factors play two roles in the bosonization procedure. It keeps
track of changes in the fermionic number and ensures that vertex operators
from different bosonic species anti-commutate\citep{delft1998bosonization}.
The fermionic number is fixed in equilibrium quantities, hence it
is possible to represent the Klein factors as simple Majoranas satisfying\citep{schulz1998fermi,senechal1999introduction}
\begin{equation}
\left\{ \eta_{\alpha},\eta_{\beta}\right\} =2\delta_{\alpha,\beta}.
\end{equation}

Defining the bosonic fields

\begin{align}
\phi & =\phi_{\uparrow}+\phi_{\downarrow},\label{eq:phi-def}\\
\theta & =\phi_{\downarrow}-\phi_{\uparrow},\label{eq:theta-def}
\end{align}
the edge Hamiltonian, $H_{\text{edge}}^{\left(j\right)}$, becomes

\begin{equation}
H_{\text{edge}}^{\left(j\right)}=\frac{v}{8\pi}\int_{-\infty}^{\infty}dx\frac{1}{g}\left(\partial_{x}\phi_{j}\right)^{2}+g\left(\partial_{x}\theta_{j}\right)^{2},\label{H0_bosonic}
\end{equation}
with the renormalized velocity \ensuremath{v} and the Luttinger parameter
\ensuremath{g} given by 
\begin{align}
v & =v_{0}\sqrt{\left(1+\frac{g_{2}}{2}\right)\left(1-\frac{g_{2}}{2}\right)},\\
g & =\sqrt{\left(1-\frac{g_{2}}{2}\right)/\left(1+\frac{g_{2}}{2}\right)}.
\end{align}
A repulsive interaction, positive $g_{2}$ in Eq.\,(\ref{eq:fermionic-interaction}),
correspond to $g<1$, while an attractive interaction yields $g>1$. 

It is tempting to start the discussion of the junction conductance
using the single electron tunneling, Eq.\,(\ref{eq:eq:fermionic-tunneling}).
However, it is known that $H_{e}$ is an irrelevant operator (suppressed
at low temperatures and long times) for $g\neq1$\citep{hou2009corner,teo2009critical,dolcetto2016edge}.
The physical reason is straightforward, the fundamental excitation
of the free bosonic theory, Eq.\,(\ref{H0_bosonic}), are define by
the primary fields $e^{\pm i\phi_{j}}$ and $e^{\pm i\theta_{j}}$\citep{difrancesco1997conformal}.
Hence, just like in the description of tunneling to/from edge states
in the quantum Hall effect\citep{wen1990chiral,wen1995topological},
the important operators are the quasi particles constructed from these
primary fields. Consequently, two correlated tunneling operators must
be added to the description of the junction,

\begin{align}
\psi_{\downarrow,j}^{\dagger}\psi_{\uparrow,j} & \sim e^{-i\phi_{j}},\label{eq:spin-flip}\\
\psi_{\downarrow,j}^{\dagger}\psi_{\uparrow,j}^{\dagger} & \sim e^{-i\theta_{j}},\label{eq:cooper-pair}
\end{align}
which correspond to the following physical processes:
\begin{itemize}
\item [${i}$:] a correlated spin-flip event,  described by
\begin{equation}
H_{s}^{\left(j,k\right)}=\lambda_{s}e^{\pm2ir\varphi}\psi_{\uparrow,k}^{\dagger}\left(0\right)\psi_{\uparrow,j}\left(0\right)\psi_{\downarrow,j}^{\dagger}\left(0\right)\psi_{\downarrow,k}\left(0\right)+\text{h.c.,}\label{eq:Hs_fermions}
\end{equation}
where spin current is transfer between edges without any net charge
current crossing the junction.
\item [${ii}$:] a correlated pair tunneling,
\begin{equation}
H_{c}^{\left(j,k\right)}=\lambda_{c}\psi_{\downarrow,j}^{\dagger}\left(0\right)\psi_{\downarrow,k}\left(0\right)\psi_{\uparrow,j}^{\dagger}\left(0\right)\psi_{\uparrow,k}\left(0\right)+\text{h.c.,}\label{eq:Hc_fermions}
\end{equation}
where a charge $2e$ crosses the junction, without spin current flow. 
\end{itemize}
Even if these operators are not initially present, they
are generated by the renormalization of $H_{e}$ in the low-energy
theory. 

After rescaling the bosonic fields, $\phi_{j}\to\phi_{j}/\sqrt{g}$
and $\theta_{j}\to\sqrt{g}\theta_{j}$ in Eq.\,(\ref{H0_bosonic})
and using the definitions from Table\,\ref{tab:Klein-factors-representation},
the three tunneling operators are

\begin{align}
H_{e} & =\sum_{k=1}^{3}2\lambda_{e}\tau^{k}\sin\left[\sqrt{g}\frac{\phi_{k+1}-\phi_{k}}{2}-\frac{\theta_{k+1}-\theta_{k}}{2\sqrt{g}}+r\varphi\right]\nonumber \\
 & +2\lambda_{e}\sigma^{k}\sin\left[\sqrt{g}\frac{\phi_{k+1}-\phi_{k}}{2}+\frac{\theta_{k+1}-\theta_{k}}{2\sqrt{g}}-r\varphi\right],\label{eq:He-jk-boson}\\
H_{s} & =\sum_{k=1}^{3}-2\lambda_{s}\tau^{k}\otimes\sigma^{k}\cos\left[\sqrt{g}\left(\phi_{k+1}\left(0\right)-\phi_{k}\left(0\right)\right)+2r\varphi\right],\label{eq:Hs-jk-boson}\\
H_{c} & =\sum_{k=1}^{3}2\lambda_{c}\tau^{k}\otimes\sigma^{k}\cos\left[\frac{\theta_{k+1}\left(0\right)-\theta_{k}\left(0\right)}{\sqrt{g}}\right],\label{eq:Hc-jk-boson}
\end{align}
where $k=\left\{ 1,2,3\right\} $ labels the edges and $k=4\to k=1$. 

It is important to highlight a technical point here. Bosonization
is a statement relating correlation functions of bosonic vertices
and fermions, hence some conclusions can only be understood when considering
expectation values of these operators. One of these statements is
that Klein factors in an equilibrium calculation will eventually produce
a plus or minus sign to an expectation value. The Klein factors in
the pair hopping events commute with each other, which might suggests
that can be neglected. However, the $Y$ junction model allows for
the expectation value of three point functions to be non zero in the
zero-temperature imaginary-time formalism, for example
\begin{align}
\left\langle H_{c}^{\left(1,3\right)}\left(\tau_{3}\right)H_{c}^{\left(3,2\right)}\left(\tau_{2}\right)H_{c}^{\left(2,1\right)}\left(\tau_{1}\right)\right\rangle _{0} & =\frac{-\lambda_{c}^{3}}{\prod_{j<k=1}^{3}\left|\tau_{k}-\tau_{j}\right|^{\frac{2}{g}}},\\
\left\langle H_{s}^{\left(1,3\right)}\left(\tau_{3}\right)H_{s}^{\left(3,2\right)}\left(\tau_{2}\right)H_{s}^{\left(2,1\right)}\left(\tau_{1}\right)\right\rangle _{0} & =\frac{+\lambda_{s}^{3}e^{+i6r\varphi}}{\prod_{j<k=1}^{3}\left|\tau_{k}-\tau_{j}\right|^{2g}}.
\end{align}
It is straightforward to show that the Klein factors in all expectations
values with an even number of vertices will produce a plus sign, however
with an odd number of vertices there must be a sequence of Pauli matrices,

\begin{equation}
\left(\tau^{1}\tau^{2}\tau^{3}\right)\otimes\left(\sigma^{1}\sigma^{2}\sigma^{3}\right)=-1,
\end{equation}
that can produce a negative sign. 

Therefore the effect of Klein factors can be effectively incorporated
as a sign change in the coupling constants in Eqs.\,(\ref{eq:Hs-jk-boson} and \ref{eq:Hc-jk-boson}),
allowing them to be safely omitted from explicit calculations.

\begin{table}
\begin{tabular}{|c|c|}
\hline 
$\eta_{\downarrow1}\eta_{\downarrow3}=-i\sigma^{1}$ & $\eta_{\uparrow1}\eta_{\uparrow3}=i\tau^{1}$\tabularnewline
\hline 
$\eta_{\downarrow2}\eta_{\downarrow1}=-i\sigma^{2}$ & $\eta_{\uparrow2}\eta_{\uparrow1}=i\tau^{2}$\tabularnewline
\hline 
$\eta_{\downarrow3}\eta_{\downarrow2}=-i\sigma^{3}$ & $\eta_{\uparrow3}\eta_{\uparrow2}=i\tau^{3}$\tabularnewline
\hline 
\end{tabular}

\caption{\label{tab:Klein-factors-representation}Product of Klein factors
represented as Pauli matrices.}
\end{table}

The rotation defined in Table \ref{tab:boson_def} leaves the free
bosonic Hamiltonian invariant,
\begin{equation}
H_{\text{free edge}}=\frac{v}{8\pi}\sum_{k=0}^{2}\int_{-\infty}^{\infty}dx\left(\partial_{x}\Phi_{k}\right)^{2}+g\left(\partial_{x}\Theta_{k}\right)^{2},\label{eq:H0_boson_new}
\end{equation}
but it enables a much more compact expression for the tunneling operators.

\begin{table}
\[
\begin{array}{|l|l|}
\hline \Phi_{0}=\frac{1}{\sqrt{3}}\left(\phi_{1}+\phi_{2}+\phi_{3}\right) & \Theta_{0}=\frac{1}{\sqrt{3}}\left(\theta_{1}+\theta_{2}+\theta_{3}\right),\\
\Phi_{1}=\frac{1}{\sqrt{2}}\left(\phi_{1}-\phi_{2}\right) & \Theta_{1}=\frac{1}{\sqrt{2}}\left(\theta_{1}-\theta_{2}\right),\\
\Phi_{2}=\frac{1}{\sqrt{6}}\left(\phi_{1}+\phi_{2}-2\phi_{3}\right) & \Theta_{2}=\frac{1}{\sqrt{6}}\left(\theta_{1}+\theta_{2}-2\theta_{3}\right),
\\\hline \end{array}
\]

\caption{\label{tab:boson_def}A linear combination of the field in the ``Y''
junctions that allows for a compact notation of the tunneling terms
in the Hamiltonian\citep{chamon2003junctions,oshikawa2006junctions}.}
\end{table}

\begin{align}
H_{e} & =\sum_{k=1}^{3}-2\lambda_{e}\tau^{k}\sin\left[\sqrt{\frac{g}{2}}\vec{b}_{k}.\vec{\Phi}\left(0\right)-\frac{\vec{b}_{k}.\vec{\Theta}\left(0\right)}{\sqrt{2g}}+r\varphi\right]\nonumber \\
 & -2\lambda_{e}\sigma^{k}\sin\left[\sqrt{\frac{g}{2}}\vec{b}_{k}.\vec{\Phi}\left(0\right)+\frac{\vec{b}_{k}.\vec{\Theta}\left(0\right)}{\sqrt{2g}}-r\varphi\right],\label{eq:He-boson}\\
H_{s} & =\sum_{k=1}^{3}2\lambda_{s}\cos\left[\sqrt{2g}\vec{b}_{k}.\vec{\Phi}\left(0\right)+2r\varphi\right],\label{eq:Hs-boson}\\
H_{c} & =\sum_{k=1}^{3}2\lambda_{c}\cos\left[\sqrt{\frac{2}{g}}\vec{b}_{k}.\vec{\Theta}\left(0\right)+\pi\right],\label{eq:Hc-boson}
\end{align}
with 
\begin{align}
\vec{\Phi} & =\left(\Phi_{1},\Phi_{2}\right),\\
\vec{\Theta} & =\left(\Theta_{1},\Theta_{2}\right).
\end{align}

The general model that I will discuss from now on is

\begin{equation}
H=H_{\text{free edge}}+H_{e}+H_{s}+H_{c}\label{eq:general_bosonic_model}
\end{equation}

\section{Scattering matrix for non-interacting fermions\label{sec:Conductance-tensor-for-g1}}

This section adopts a known result from the literature \citep{oshikawa2006junctions}.
I will review it briefly to make the discussion reasonably self-contained
and to establish the notation.

For non-interacting systems, the first step in evaluating the conductance
tensor within linear response is to determine the scattering matrix
for the Y-junction\citep{chen2002landauertype,egger2003transport,oshikawa2006junctions,giuliano2022multiparticle}. 

Each spin projection in the Hamiltonian $H_{\text{free edge}}^{\left(j\right)}$ 
violates parity\citep{difrancesco1997conformal} (left/right reflection)  and demands careful handling.
A discrete \emph{naive} version for each spin projection of $H_{\text{free edge}}^{\left(j\right)}$ can
be written as

\begin{equation}
H_{L}=i\sum_{j=1}^{3}\sum_{n=\infty}^{\infty}c_{n,j}^{\dagger}\left(c_{n+1,j}-c_{n,j}\right),\label{eq:discrete-edge}
\end{equation}
where the continuum limit is defined via $\psi\left(x=na\right)=\lim_{a\to0}c_{n}/\sqrt{a}$.
However,  $H_{L}$ is not Hermitian and the absence of parity symmetry (reflection around the $n=0$ site) in this formulation obscures
the expression for the tunneling Hamiltonian, $H_{e}^{\left(j,k\right)}$,
in terms of the discrete $c-$fermions from Eq.\,(\ref{eq:discrete-edge}).
A simple tunneling term between edges using the $c$-fermions will generate
a scattering matrix that does not conserve probability. 

The standard procedure in the literature to recover Hermiticity and to impose parity symmetry
involves first doubling the degrees of freedom through the inclusion
of right-moving fermions\citep{difrancesco1997conformal},

\begin{align}
H_{L}+H_{L}^{\dagger} & =i\sum_{j=1}^{3}\sum_{n=-\infty}^{\infty}c_{n,j}^{\dagger}\left(c_{n+1,j}-c_{n-1,j}\right),
\end{align}
To recover the correct number of physical degrees of freedom, the
sum is truncated at $n=0$, yielding the effective Hamiltonian

\begin{equation}
\hat{H}_{0}=\sum_{j=1}^{3}\sum_{n=1}^{\infty}ic_{n,j}^{\dagger}\left(c_{n+1,j}-c_{n-1,j}\right)+ic_{0,j}^{\dagger}c_{1,j}.\label{eq:H0_Neumann}
\end{equation}
A final step involves a gauge transformation of the fermionic operators,
$d_{n,j}=e^{i\frac{\pi}{2}n}c_{n,j}$, which maps the model onto a
conventional tight-binding defined on a half-line,

\begin{align}
\hat{H}_{0} & =\sum_{j=1}^{3}\sum_{n=1}^{\infty}d_{n,j}^{\dagger}\left(d_{n+1,j}+d_{n-1,j}\right)+d_{0,j}^{\dagger}d_{1,j}.\label{half-chain}
\end{align}
An equivalent discussion, but starting from Eq.~(\ref{half-chain}) and arriving at the continuous version of Eq.(\ref{eq:discrete-edge}), can be found in appendixes A.1 and A.2 of  \citet{oshikawa2006junctions}.

Now that the discrete version of the edge model is Hermitian and preserves parity,
it is correct to write the tunneling between edges as 

\[
H_{\tilde{t}}=\sum_{j=1}^{3}\tilde{t}e^{i\tilde{\varphi}}d_{0,j+1}^{\dagger}d_{0,j}+\text{h.c.}.
\]

This entire procedure is known as ``folding'' a chiral fermion chain
into a tight-biding half-chain\citep{eggert1992magnetic}. It is important
to note that in these new fermions the chemical potential was shifted
from momentum $k=0$ to $k=\pm\frac{\pi}{2}$. Furthermore, right
moving fermions correspond to the original left moving fermions after
passing through the junction position $\left(j=0\right)$. 

The scattering matrix for $\hat{H}_{0}+H_{\tilde{t}}$ was found in
\citet{oshikawa2006junctions}, and here I outline their solution.
The diagonalization procedure follows the same steps from solving
the Bogoliubov-de Gennes equations. For a site $n$ away from the
boundary, the coefficients of linear transformation that diagonalize
the Hamiltonian must satisfy the wave equation, 

\begin{equation}
\alpha_{n-1,j}+\alpha_{n+1,j}=\Lambda\alpha_{n,j}
\end{equation}
while at the boundary they must satisfy

\begin{align}
\alpha_{1,j}+\tilde{t}e^{i\tilde{\varphi}}\alpha_{0,j+1}+\tilde{t}e^{-i\tilde{\varphi}}\alpha_{0,j-1} & =\Lambda\alpha_{0,j}.
\end{align}

Applying the Ansatz

\begin{equation}
\alpha_{n,j}=L_{j}e^{-ikn}+R_{j}e^{ikn},
\end{equation}
and solving for the scattering amplitudes, $\vec{R}=S\vec{L}$, gives
the same scattering matrix previously derived for the Y-junction of
three quantum wires\citep{oshikawa2006junctions},

\begin{equation}
S=-e^{2ik}\frac{1-\tilde{t}e^{-ik}M}{1-\tilde{t}e^{ik}M},
\end{equation}
where $M=e^{i\tilde{\varphi}}\Omega+e^{-i\tilde{\varphi}}\Omega^{-1}$,
and 
\begin{equation}
\Omega=\left[\begin{array}{ccc}
0 & 1 & 0\\
0 & 0 & 1\\
1 & 0 & 0
\end{array}\right].
\end{equation}

The scattering matrix can be explicitly written as

\begin{equation}
S\left(\tilde{t},k,\varphi\right)=e^{2ik}\left[S_{0}+S_{+}\Omega+S_{-}\Omega^{-1}\right],
\end{equation}
where

\begin{align}
S_{0} & =\frac{1}{D}\left[2\tilde{t}^{3}e^{ik}\cos\left(3\tilde{\varphi}\right)+\tilde{t}^{2}\left(e^{2ik}+2\right)-1\right],\\
S_{+} & =\frac{\tilde{t}e^{i\tilde{\varphi}}}{D}\left[\left(e^{-ik}-e^{ik}\right)+\left(1-e^{2ik}\right)\tilde{t}e^{-3i\tilde{\varphi}}\right],\\
S_{-} & =\frac{\tilde{t}e^{-i\tilde{\varphi}}}{D}\left[\left(e^{-ik}-e^{ik}\right)+\left(1-e^{2ik}\right)\tilde{t}e^{3i\tilde{\varphi}}\right],\\
D & =1-3\tilde{t}^{2}e^{2ik}-2\tilde{t}^{3}e^{3ik}\cos\left(3\tilde{\varphi}\right)
\end{align}

As expected the noninteracting problem has a continuous dependence
on $\tilde{t}$ and $k$. In the renormalization group language, this
means that the tunneling operator is an exactly marginal operator
and there is a continuous line of renormalization group fixed points
as a function of $\tilde{t}$. Once interactions are introduced, in
general, there will be a renormalization group flow. Nevertheless,
this exactly solvable noninteracting case provides a useful framework
for the identifying important fixed points of the interacting theory\citep{oshikawa2006junctions}.

\begin{figure}

\includegraphics[width=0.75\columnwidth]{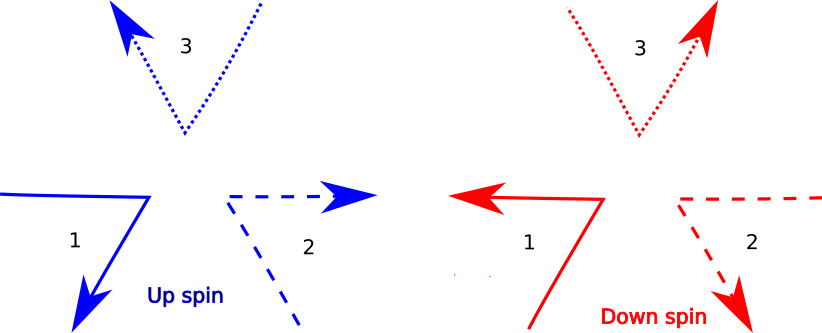}\caption{\label{fig:The-zero-conductance}The zero conductance scattering matrix,
$\Gamma_{0}$. }

\end{figure}

\begin{figure}

\includegraphics[width=0.75\columnwidth]{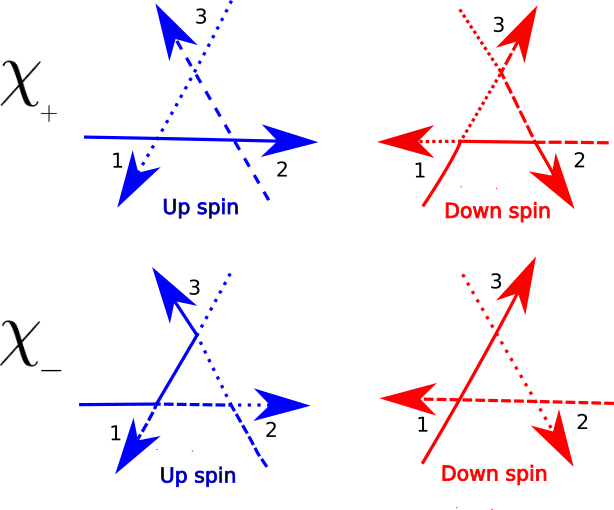}\caption{\label{fig:The-Xi_plus-scattering}For a given $\pm\tilde{\varphi}$,
the up spin channel will have a $\chi_{\pm}$ scattering matrix, while
the down spin channel will have a $\chi_{\mp}$ scattering matrix.
The line type indicates the edge that the fermion originates, solid
for edge $1$, dashed for edge $2$ and dotted for edge $3$.}
\end{figure}

\begin{itemize}
\item [{$\Gamma_{0}$}:] $\tilde{t}\to0,$$S=-e^{2ik}$.\\
A perfect reflection on the folded line with left/right-moving fermions
corresponds to no scattering in the original left moving fermions
(see Fig \ref{fig:The-zero-conductance}).
\item [{$\Gamma_{A}$}:] $\tilde{t}\to\infty,$$S=-I$.\\
A left-moving fermion with momentum $k$ is reflected into a right
moving hole with momentum $k$. This corresponds to an Andreev's reflection
event at the junction, a charge $2e$ leaves/enters a terminal of
the junction, and the scattering matrix does not depend on $\varphi$.
\item [{$\Gamma_{M}$}:] $\left|S_{+}\right|=\left|S_{-}\right|$.\\
A fermion arriving at the junction on edge $j$ has equal probability
to tunnel to edge $j+1$ and $j-1$. 
Most of the properties of this fixed point are unknown due to the
non trivial effect of the Klein factors in the problem with interactions\citep{chamon2003junctions,oshikawa2006junctions}.
\item [{$\chi_{+,\uparrow/\downarrow}$}:] $\left\{ \tilde{t}\to1,k=-3\tilde{\varphi}+\left(2n+1\right)\pi\right\} ,\left\{ S_{0}=S_{-}=0\right\}$.\\
This particular set of parameters allows for a perfect chiral scattering.
A left-moving fermion arriving at the junction on edge $j$ has unity
probability to tunnel to edge $j+1$ (see Fig \ref{fig:The-Xi_plus-scattering}).
This fixed point explicitly breaks time reversal symmetry, hence it
will not be present in a $Y$ junction made of edge states.
\item [{$\chi_{-,\uparrow/\downarrow}$}:] $\left\{ \tilde{t}\to1,k=3\tilde{\varphi}+\left(2n+1\right)\pi\right\} ,\left\{ S_{0}=S_{+}=0\right\}$.\\
Similar to the previous case, a left-moving fermion arriving at the
junction on edge $j$ has unity probability to tunnel to edge $j-1$. 
\end{itemize}

\section{stability of the weak tunneling fixed point\label{sec:stability-of-the}}

In the limit $\lambda_{e,s,c}=0$ the fermions are moving without
any scattering at the position of the junction, hence $\left\{ \Phi_{i}\left(0,t\right),\Theta_{i}\left(0,t\right)\right\} $
exhibit Gaussian fluctuations governed by Eq.\,(\ref{eq:H0_boson_new}).
A ``weak'' coupling analyses of Eq.\,(\ref{eq:general_bosonic_model})
corresponds to introduce $0<\lambda_{e,s,c}\ll1$ and studying the
stability of the Gaussian theory\citep{gogolin2004bosonization}. 

The scaling dimension of the tunneling operators around the Gaussian
fixed point can be readily read,

\begin{align}
\frac{\partial\lambda_{e}}{\partial\ell} & =\left(1-\frac{1}{2}\left(g+\frac{1}{g}\right)\right)\lambda_{e},\label{eq:lambda_e}\\
\frac{\partial\lambda_{s}}{\partial\ell} & =\left(1-2g\right)\lambda_{s},\\
\frac{\partial\lambda_{c}}{\partial\ell} & =\left(1-\frac{2}{g}\right)\lambda_{c}.
\end{align}
These equations were also derived in the context of a single point
contact model\citep{hou2009corner,teo2009critical,dolcetto2016edge}.
They show that the single electron tunneling is \emph{always} an irrelevant
perturbation. For the range $1/2<g<2$, the correlated hopping amplitudes
also renormalize to zero $\left(\lambda_{s,c}\to0\right)$. Hence,
in the zero temperature limit, $T\to0$, no current flows across the
junction in linear response, characterizing the zero conductance fixed
point $\Gamma_{0}$.

In this regime and at finite temperatures, the conductance can be
evaluated using perturbation theory improved by renormalization group.
The finite temperature corrections to the conductance are power laws
determined by the scaling dimensions of the corresponding leading
irrelevant operators\citep{teo2009critical}, and as a function of
$g$ they are given by
\begin{align}
G_{e} & \propto T^{\left(g+\frac{1}{g}\right)-2}\,\text{for \ensuremath{\frac{1}{\sqrt{3}}<g<\sqrt{3}}},\label{eq:finte-temp-1}\\
G_{s} & \propto T^{4g-2}\,\text{for \ensuremath{\frac{1}{2}<g<\frac{1}{\sqrt{3}}}},\label{eq:finte-temp-2}\\
G_{c} & \propto T^{\frac{4}{g}-2}\,\text{for \ensuremath{\sqrt{3}<g<2}}.\label{eq:finte-temp-3}
\end{align}

The zero conductance fixed point, $\Gamma_{0}$, is unstable in the
following regimes:
\begin{itemize}
\item [${i}$:] for $g<1/2$, unstable to $H_{s}$ ; 
\item [${ii}$:] for $g>2$, unstable to $H_{c}$. 
\end{itemize}
Because the single particle hopping is always irrelevant, the $Y$
junction dynamics is described by 

\begin{equation}
H=H_{\text{free edge}}+H_{s,c},\label{relevantH}
\end{equation}
in these two regimes.

The $g=1/2$ and $g=2$ cases are marginal and can be mapped into
a non-interacting fermionic theory. In the next section I will discuss
these two marginal cases, leveraging the results from the $g=1$ scattering
matrix, before addressing the two relevant renormalization group flows.

\section{\label{sec:refermionization-at-2}refermionization at $g=1/2$ and
$2$}

The pair tunneling operators, $H_{s,c}$, have scaling dimension one
at $g=1/2$ and $2$. Hence, it is possible to refermionize the theory
to a new set of non-interacting fermionic fields\citep{delft1998bosonization,gogolin2004bosonization}.
Since the procedure is the same to both cases, I introduce a phase
$\hat{\varphi}$, defined as zero for $H_{c}$ and $2r\varphi$ for
$H_{s}$, and treat both cases simultaneously.

At $g=\left\{ \frac{1}{2},2\right\}$ the renormalization group analysis
around the Gaussian fixed point shows that at low energies, only the
Hamiltonian Eq.~(\ref{relevantH}) needs to be considered.

There are several ways to proceed, but a particularly insightful way
is to consider the zero-temperature partition function,

\begin{equation}
Z=\lim_{\beta\to\infty}e^{-\beta H},
\end{equation}
by expanding it to all orders in $\lambda=\lambda_{\left\{ s,c\right\} }$,
and then exactly integrating out the free bosonic fields. This procedure
yields a partition function identical to a one-dimensional Coulomb
gas comprise of particles with fractional ``electric'' charges,

\begin{align*}
\vec{e}_{\alpha=1,23} & =\begin{cases}
\left(1,0\right)\\
\left(-\frac{1}{2},\frac{\sqrt{3}}{2}\right)\\
\left(-\frac{1}{2},-\frac{\sqrt{3}}{2}\right)
\end{cases}
\end{align*}
that interact via a logarithm interaction
\[
\ln\left|\tau_{i}-\tau_{j}\right|\vec{e}_{\alpha}\left(\tau_{i}\right).\vec{e}_{\beta}\left(\tau_{j}\right),
\]
and have fugacity $y\propto\lambda e^{\pm i\hat{\varphi}}$. Only
charge-neutral configurations, satisfying 
\begin{equation}
\sum_{i}\vec{e}_{\alpha}\left(\tau_{i}\right)=0,
\end{equation}
contribute to the partition function.

The same Coulomb gas is also obtained by considering the bosonic partition
function 
\begin{equation}
Z=\lim_{\beta\to\infty}e^{-\beta\tilde{H}},
\end{equation}
where
\begin{align}
\tilde{H} & =v\sum_{j=1}^{3}\int_{-\infty}^{\infty}dx\left(\partial\tilde{\phi}_{j}\right)^{2}\nonumber \\
 & +\lambda e^{i\hat{\varphi}}\tau^{k}\otimes\sigma^{k}e^{i\left[\tilde{\phi}_{j+1}\left(0\right)-\tilde{\phi}_{j}\left(0\right)\right]}+\text{h.c..}\label{eq:Y-bosons_c}
\end{align}
The final step is to use the bosonization identity 
\begin{equation}
\tilde{\psi}_{j}\left(0\right)\sim\eta_{\uparrow j}e^{-i\tilde{\phi}_{j}\left(0\right)},\label{eq:refermionization-1}
\end{equation}
and represent the remaining Klein factors using the Pauli matrices,
Tab.\,\ref{tab:Klein-factors-representation}.

Hence, the original partition function with $g=\left\{ \frac{1}{2},2\right\} $
is equivalently rewritten as a quadratic fermionic model

\begin{align}
\tilde{H} & =iv\sum_{j=1}^{3}\int_{-\infty}^{\infty}dx\tilde{\psi}_{j}\partial_{x}\tilde{\psi}_{j}\nonumber \\
 & +i\lambda e^{i\hat{\varphi}}\sigma^{j}\tilde{\psi}_{j+1}^{\dagger}\left(0\right)\tilde{\psi}_{j}\left(0\right)+\text{h.c.}.\label{eq:refermionizationH}
\end{align}
It is important to emphasize that the $\tilde{\psi}_{j}'s$ are not
the original fermions; rather, they correspond to a pair tunneling,
Eqs.\,(\ref{eq:spin-flip}-\ref{eq:cooper-pair}). Consequently, there
are only three fermionic fields instead of the original six.

Following the same procedure that was discussed for $g=1$, the scattering
matrix for $g=2$ and $1/2$ is

\begin{equation}
S=-e^{2ik}\frac{1-\lambda e^{-ik}\tilde{M}}{1-\lambda e^{ik}\tilde{M}},\label{eq:scattering-interactions}
\end{equation}
where $M=e^{i\hat{\varphi}}\tilde{\Omega}+e^{-i\hat{\varphi}}\tilde{\Omega}^{-1}$,
and 
\begin{equation}
\tilde{\Omega}=\left[\begin{array}{ccc}
0 & -i\sigma^{2} & 0\\
0 & 0 & -i\sigma^{3}\\
-i\sigma^{1} & 0 & 0
\end{array}\right].
\end{equation}
 The scattering matrix can once again be parametrize as 
\begin{equation}
S\left(\lambda_{c,s},k,\hat{\varphi}\right)=-e^{2ik}\left(S_{0}+S_{+}\tilde{\Omega}+S_{-}\tilde{\Omega}^{-1}\right),
\end{equation}
 with

\begin{align}
S_{0} & =\frac{1-\lambda^{2}e^{2ik}-2\lambda^{2}+2\lambda^{3}e^{ik}\cos\left(3\hat{\varphi}\right)}{D},\\
S_{+} & =\frac{\lambda e^{i\hat{\varphi}}\left[\left(e^{ik}-e^{-ik}\right)\left(1-\lambda e^{ik-3i\hat{\varphi}}\right)\right]}{D},\\
S_{-} & =\frac{\lambda e^{-i\hat{\varphi}}\left[\left(e^{ik}-e^{-ik}\right)\left(1-\lambda e^{ik+3i\hat{\varphi}}\right)\right]}{D},\\
D & =\left(1-\lambda^{2}e^{2ik}\right)-2\lambda^{2}e^{3ik}\left(1-\lambda\cos\left(3\hat{\varphi}\right)\right).
\end{align}

Again the same fixed points are presents in the scattering matrix:
\begin{itemize}
\item [{$\Gamma_{0}$:}] no current across the junction. 
\item [{$\Gamma_{A}$:}] the ``Andreev'' like fixed point again corresponds
to $\lambda=\lambda_{c,s}\to\infty$, it can be interpreted as the
tunneling of 4 of the original fermions. Chiral symmetry is restored
in the transport processes across the junction, since the scattering
matrix is $\hat{\varphi}$ independent.
\item [{$\Gamma_{M}$:}] for $\hat{\varphi}=\left\{ 0,\pi\right\} $ the
scattering matrix has $\left|S_{+}\right|=\left|S_{-}\right|$ for
all value of $\lambda$.
\end{itemize}

\section{strong tunneling fixed points: Duality and Quantum Brownian motion\label{sec:strong-tunneling-fixed}}

The standard approach to analyzing the strong tunneling fixed point,
$\Gamma_{A}$, is based on duality, where the starting point is the
partition function

\begin{equation}
Z_{s,c}=\lim_{\beta\to\infty}e^{-\beta\left(H_{\text{free edge}}+H_{s,c}\right)},
\end{equation}
where the subscripts denote the different tunneling processes. 

In the limit $\lambda_{s,c}\to\infty$ the $\vec{\Phi}/\vec{\Theta}$
fields become pinned to the values that minimize the periodic potentials
in Eqs.\,(\ref{eq:Hs-boson}-\ref{eq:Hc-boson}). Due to the structure
of these potentials, the minima form a lattice in field space. Understanding
the low-energy dynamics in this regime requires considering \emph{instanton
events} (quantum transitions in which the system tunnels between adjacent
minima). Analyzing these instantons leads to a dual theory, in which
tunneling events become the main dynamical processes.

\begin{figure}
\includegraphics[trim={50px 50px 50px 50px},clip,width=1\columnwidth]{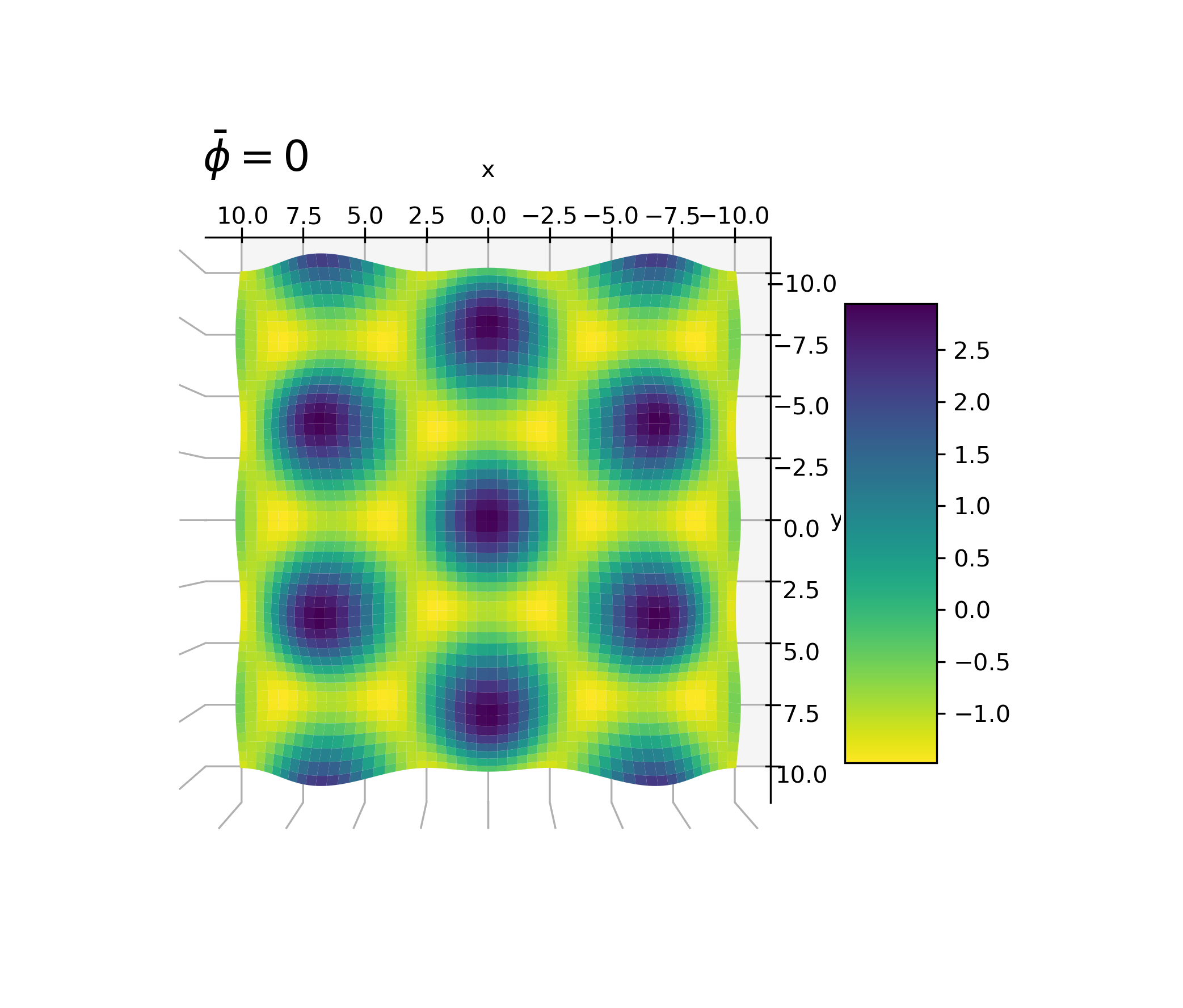}

\includegraphics[trim={50px 50px 50px 50px},clip,width=1\columnwidth]{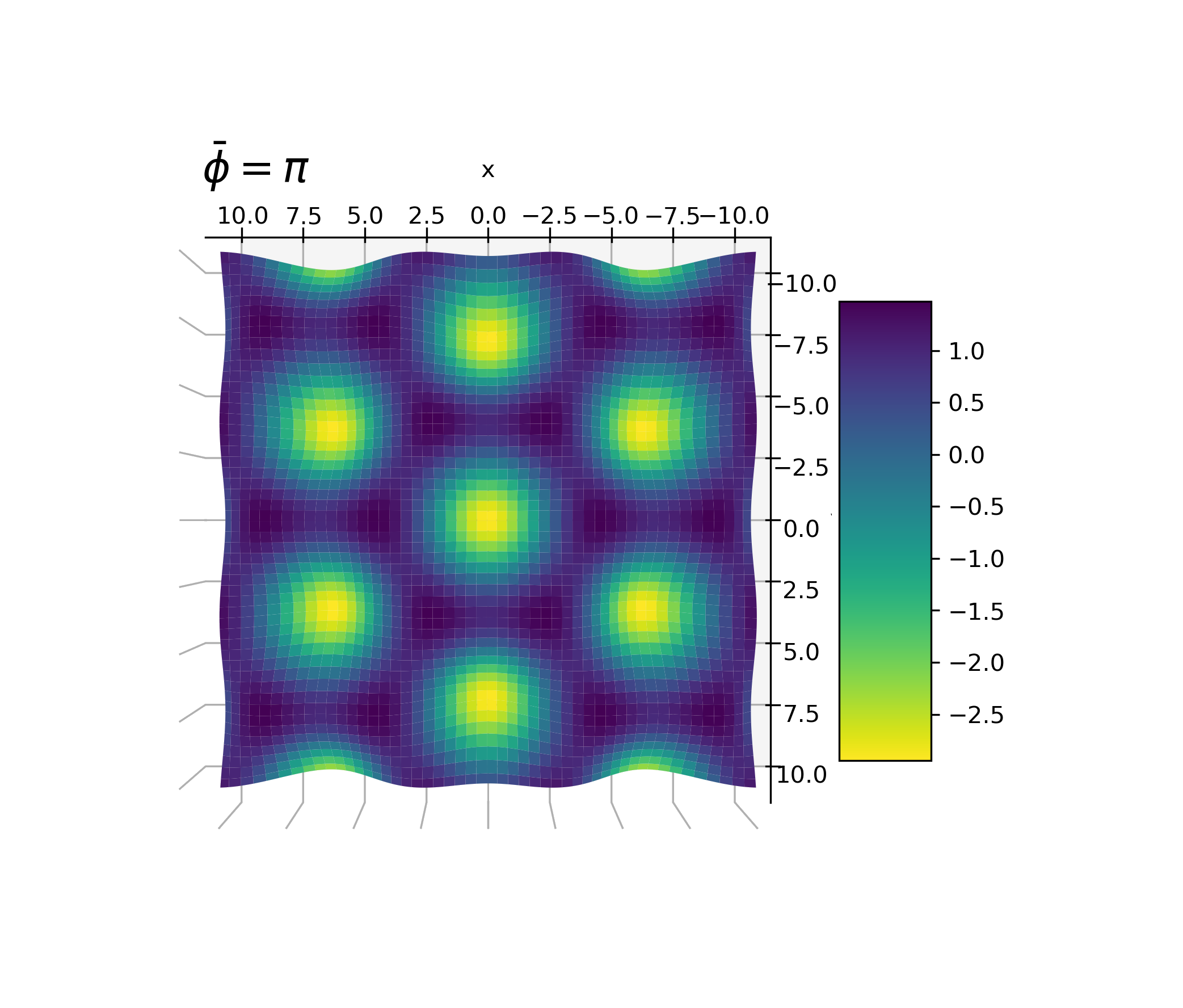}

\caption{\label{fig:lattice-minima}A plot of the function $F$ for $\bar{\varphi}=0$
and $\pi$.}
\end{figure}

In the case of a point contact tunneling\citep{kane1992transmission,kane1992transport,eggert1992magnetic}
the dual theory is equivalent to the quantum Brownian motion of a
particle in a one-dimensional periodic potential\citep{caldeira1983quantum,guinea1985dynamics}.
The mobility $\mu$ of the particle can be related to the conductance
in the transport problem\citep{oshikawa2006junctions}. This framework
generalizes naturally to the Y-junction, where the dual theory describes
a particle undergoing Brownian motion on a \emph{two-dimensional triangular
lattice}\citep{yi1996quantum,affleck2001quantum,yi2002resonant,chamon2003junctions,oshikawa2006junctions}.
This analogy provides a powerful physical picture of the renormalization
group (RG) fixed points:
\begin{itemize}
\item [{${i}$:}] $\Gamma_{0}$ is the weak tunneling fixed point. It corresponds to
a \emph{delocalized particle}, with equal probability of occupying
any potential minimum. This results in full mobility, $\mu=1$, of
the Brownian particle, corresponding to zero conductance in the $Y$
junction transport problem.
\item [{${ii}$:}] $\Gamma_{A}$ is the strong tunneling fixed point. It corresponds to
a \emph{localized particle}, confined to a single potential well.
This leads to zero mobility, $\mu=0$. 
In the case of a Luttinger liquid interfacing
with a superconductor, this corresponds to an enhanced conductance
due to Andreev reflection. 
\item [{${iii}$:}] Intermediate fixed points describe nontrivial Brownian dynamics, where
the particle exhibits \emph{partial mobility}, resulting in a nontrivial
conductance.
\end{itemize}
The classical analysis for the periodic potentials in Eqs.\,(\ref{eq:Hs-boson} and \ref{eq:Hc-boson})
corresponds to minimizing the function

\begin{equation}
F=\sum_{j=1}^{3}\cos\left[\vec{b}_{j}.\left(x,y\right)+\bar{\varphi}\right].\label{eq:Ffunction}
\end{equation}
Since this potential is periodic, the minima form a lattice consistent
with the reciprocal lattice structure,

\begin{equation}
\left(x,y\right)=\frac{4\pi}{\sqrt{3}}n\vec{a}_{1}+\frac{4\pi}{\sqrt{3}}m\vec{a}_{2}+\vec{\ell},
\end{equation}
with $\left(n,m\right)$ integers. There are three sublattices that
are local extrema for $F$,

\begin{align}
\vec{\ell}_{A} & =\left(0,0\right)\\
\vec{\ell}_{B} & =-\frac{4\pi}{3}\vec{b}_{3}\\
\vec{\ell}_{C} & =-\frac{8\pi}{3}\vec{b}_{3}
\end{align}
with local energies

\begin{align}
F_{A} & =3\cos\left(\bar{\varphi}\right)\label{eq:FA}\\
F_{B} & =\cos\left(\frac{2\pi}{3}+\bar{\varphi}\right)-2\cos\left(\bar{\varphi}-\frac{\pi}{3}\right)\label{eq:FB}\\
F_{C} & =\cos\left(\frac{4\pi}{3}+\bar{\varphi}\right)+2\cos\left(\bar{\varphi}-\frac{2\pi}{3}\right)\label{eq:FC}
\end{align}
Hence, the absolute minima fall into two categories:
\begin{itemize}
\item  [{${i}$:}] for $\bar{\varphi}\not\equiv\left\{ 0,\frac{2\pi}{3},\frac{4\pi}{3}\right\} \text{\,mod \ensuremath{2\pi}}$,
the field values that minimize $F$, Eq.\,(\ref{eq:Ffunction}), form
a \emph{triangular lattice} with $d=\frac{4\pi}{\sqrt{3}}$ as the
smallest separation between the minima.sublattice corresponding to
the absolute minimum depends on $\bar{\varphi}$; for instance, at
$\bar{\varphi}=\pi$ the minimum is at $F_{A}$.

\item  [{${ii}$:}] for $\bar{\varphi}\equiv\left\{ 0,\frac{2\pi}{3},\frac{4\pi}{3}\right\} \text{ mod \ensuremath{2\pi}}$,
the minima form a \emph{hexagonal lattice}, since two of the sublattices
have degenerate energies. In this case, the separation between each
minima is $d=\frac{4\pi}{3}$ (see Fig.\,\ref{fig:lattice-minima}).
For example, when $\bar{\varphi}=0$, the two degenerated sublattices
correspond to $F_{B}$ and $F_{C}$.
\end{itemize}
The sign of the couplings $\lambda_{s,c}$ plays an important role,
as it corresponds to a $\pi$ phase shift. In principle $\lambda_{s,c}$
can be negative; however, a second order perturbation theory on $H_{e}$
suggests that $\lambda_{c,s}$ should be positive. From this point
onward, I \emph{assume} $\lambda_{s,c}>0$.

To improve clarity, in the following subsections, all operators written
in the\emph{ dual} \emph{theory}, that is near $\Gamma_{A}$ fixed
point, will be denoted by a \emph{bar}.

\subsection{The $\Theta$ field\label{subsec:Theta-strong}}

At weak tunneling and for $g>2$, $H_{c}$ has a relevant renormalization
group flow. The minima of $F$ lie on the $F_{A}$ triangular lattice,
\begin{equation}
\left(x,y\right)=n\vec{c}_{1}+m\vec{c}_{2},
\end{equation}
where $\vec{c}_i = \frac{4\pi}{\sqrt{3}}\sqrt{\frac{g}{2}}\vec{a}_i$.
The dual theory is described by

\begin{equation}
\bar{H}_{c}=-\bar{\lambda}_{c}\sum_{k=1}^{3}\cos\left[\sqrt{\frac{2g}{3}}\vec{a}_{k}.\vec{\Phi}\left(0\right)\right],\label{eq:Hc-dual}
\end{equation}
with $\bar{\lambda}_{c}>0$. The corresponding renormalization group
equation is

\begin{equation}
\frac{\partial\bar{\lambda}_{c}}{\partial\ell}=\left(1-\frac{2g}{3}\right)\bar{\lambda}_{c}.\label{eq:Lambda-rg}
\end{equation}
This implies that the $\Gamma_{A}$ fixed point is stable for $g>\frac{3}{2}$.

\subsubsection{The $\Gamma_{A}$ fixed point}

There is a simple physical picture to understand the infinite coupling
fixed point. When $\Gamma_{A}$ is stable the bosonic fields are pin
at a minimum. In particular, for $H_{c}$, it is pin at the $F_{A}$
sublattice, hence

\begin{equation}
\vec{\Theta}\left(0,0\right)=\left(0,0\right)\ \text{mod}\:2\pi.
\end{equation}
This means that the bosonic fields at the junction position are locked
together, 
\begin{equation}
\theta_{1}\left(0\right)=\theta_{2}\left(0\right)=\theta_{3}\left(0\right).
\end{equation}
Thus, a bosonic excitation on one of the edges is perfectly transmitted
to the others.

\subsubsection{The intermediate (unstable) fixed point}

An interesting situation arises in the region $\frac{3}{2}<g<2$,
where both the weak tunneling, $\Gamma_{0}$, and the strong tunneling
fixed point, $\Gamma_{A}$, are stable. The standard assumption in
such cases is that there is only one intermediate unstable fixed point
at a finite coupling. Hence, there would be a critical tunneling amplitude,
$\lambda_{c}^{\text{critical}}\left(g\right)$, that separates the
RG flows: for $\lambda_{c}<\lambda_{c}^{\text{critical}}$, the system
flows to $\Gamma_{0}$; otherwise, it flows to $\Gamma_{A}$. 

There are two important values of $g$. At $g=\sqrt{3}$ the theory
becomes self dual. Notably, this is also the value at which the scaling
dimensions of the pair tunneling and single-electron tunneling operators
become equal, signaling a qualitative shift in behavior. Finally,
at $g=\frac{3}{2}$ the dual theory is marginal.

To identify the nature of the intermediate unstable fixed point, I
adopt the discussion from \citealt{affleck2001quantum}. In the $\Gamma_{A}$
fixed point the field $\vec{\Theta}$ assume values that form a triangular
lattice. The dominant quantum fluctuations in this regime are instantons,
which take the field to a nearest-neighbor minima, although less
probable fluctuations to next-nearest neighbors also occur. These
next-nearest neighbors hopping
select one of three triangular sublattices of the original
lattice, minimally breaking the symmetry of the dual model. While
such quantum fluctuations are described by operators that are more irrelevant
than $\bar{H}_{c}$ near the $\Gamma_{A}$ fixed point, their duals
are relevant near $\Gamma_{0}$. For example, consider the operator

\begin{equation}
H_{\text{Potts}}=-h\sum_{k=1}^{3}\cos\left[\sqrt{\frac{2}{3g}}\vec{a}_{k}.\vec{\Theta}\left(0\right)\right].\label{eq:Potts1}
\end{equation}
with $h>0$. This operator destabilizes the $\Gamma_{0}$ fixed point
for $g>\frac{2}{3}$, but its dual is irrelevant. Similarly, the dual
operator

\begin{equation}
\bar{H}_{\text{Potts}}=-\bar{h}\sum_{k=1}^{3}\cos\left[\sqrt{\frac{2g}{9}}\vec{b}_{k}.\vec{\Phi}\left(0\right)\right]\label{eq:Potts2}
\end{equation}
destabilizes the $\Gamma_{A}$ fixed point, but its dual is irrelevant
near $\Gamma_{0}$. From these two cases, it is reasonable that $\left\{ h,\bar{h}\right\} =0$
defines the unstable intermediate fixed point. 

\citealt{affleck2001quantum} demonstrated that such theories are
connected, via conformal embedding, to the three-state Potts model
when $g=1$. The operator $H_{\text{Potts}}$ maps to transverse boundary
field in the Potts model and $\bar{H}_{\text{Potts}}$ to a longitudinal
field. This mapping naturally associates the intermediate unstable
fixed point with the universality class of the $A\,+\,B\,+\,C$ phase
of the Potts model, characterized by vanishing transverse and longitudinal
boundary fields.

\subsection{The $\Phi$ field}

The RG flow of the $\Phi$ field is towards strong tunneling in the
regime $g<\frac{1}{2}$. The correlated spin-flip tunneling, Eq.(\ref{eq:Hs-boson}),
is sensitive to the Kane-Mele phase from the microscopic model. There
are two distinct situations to be consider.

\subsubsection{$2r\varphi\protect\not\equiv\left\{ 0,\frac{2\pi}{3},\frac{4\pi}{3}\right\} \text{\,mod\,}2\pi$}

In this case, the minima of $H_{s}$ again form a triangular lattice,
with lattice spacing of $d=\frac{4\pi}{\sqrt{3}}\frac{1}{\sqrt{2g}}$.
The corresponding dual theory is given by

\begin{equation}
\bar{H}_{s}=\bar{\lambda}_{s}\sum_{k=1}^{3}\cos\left[\sqrt{\frac{2}{3g}}\vec{a}_{k}.\vec{\Theta}\left(0\right)+2r\varphi\right],
\end{equation}
with $\bar{\lambda}_{s}>0$ and the renormalization equation,

\begin{equation}
\frac{\partial\bar{\lambda}_{s}}{\partial\ell}=\left(1-\frac{2}{3g}\right)\bar{\lambda}_{s}.
\end{equation}

There are some similarities to the case with attractive interactions
(the $\Theta$-field analysis):
\begin{itemize}
\item  [{${i}$:}] there is an unstable intermediate fixed point between $\frac{1}{2}<g<\frac{2}{3}$;
\item  [{${ii}$:}] the theory is self dual at $g=\frac{\sqrt{3}}{3}$;
\item  [{${iii}$:}] the dual theory is marginal at $g=\frac{2}{3}$. 
\end{itemize}
It is evident from Eqs.\,(\ref{eq:FA}-\ref{eq:FC}) that $2r\varphi=\left\{ \pm\frac{\pi}{3},\pi\right\} \text{\,mod\,}2\pi$
are identical to the problem discussed in Sec.\,\ref{subsec:Theta-strong}.
Hence, the corresponding unstable fixed point is also related by conformal
embedding to the $A+B+C$ phase of the boundary Potts model. 

For other values of $2r\varphi$, the same general structure also
persists. However, as $2r\varphi$ approaches $\left\{ 0,\frac{2\pi}{3},\frac{4\pi}{3}\right\} \text{\,mod\,}2\pi$
the energy difference between the sublattices diminishes. As a result,
there must be a temperature-dependent cross over from the physics
of $2r\varphi=\left\{ 0,\frac{2\pi}{3},\frac{4\pi}{3}\right\} \text{\,mod\,}2\pi$
to the physics of $2r\varphi=\left\{ \pm\frac{\pi}{3},\pi\right\} \text{\,mod\,}2\pi$.

\begin{figure}
\includegraphics[width=0.9\columnwidth]{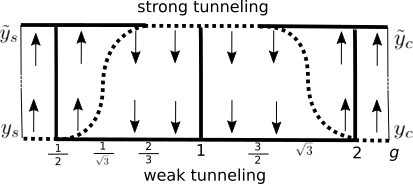}

\caption{\label{fig:Renormalization-group-diagram}Renormalization group diagram
for $\lambda_{s,c}>0$ and $r\varphi\protect\neq2\pi n$ . Solid lines
are stable and dotted lines are unstable fixed points for the renormalization
group flow. The arrows indicate the renormalization flow. At $g=\frac{1}{2},1,2$
the model is marginal and there are lines of fixed points. At $g=\left\{ \sqrt{3},\frac{1}{\sqrt{3}}\right\} $
there are changes in the temperature dependence of the conductance
around the stable weak tunneling fixed point.}
\end{figure}

A summary of the $\Theta$ and $\Phi$ flows as a function of $g$
is presented in Fig.\,\ref{fig:Renormalization-group-diagram}.

\subsubsection{$2r\varphi\equiv\left\{ 0,\frac{2\pi}{3},\frac{4\pi}{3}\right\} \text{\,mod\,}2\pi$\label{subsec:gamma_M}}

\begin{figure}
\includegraphics[width=0.9\columnwidth]{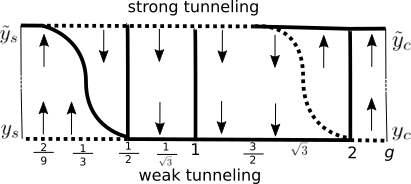}

\caption{\label{fig:Renormalization-group-diagram.}Renormalization group diagram
for $\lambda_{s,c}>0$ and $2r\varphi\equiv\left\{ 0,\frac{2\pi}{3},\frac{4\pi}{3}\right\} \text{\,mod\,}2\pi$.
Solid lines are stable and dotted lines are unstable fixed points
for the renormalization group flow. The arrows indicate the renormalization
flow. At $g=\frac{1}{2},1,2$ the model is marginal and there are
lines of fixed points. At $g=\left\{ \sqrt{3},\frac{1}{\sqrt{3}}\right\} $
there are changes in the temperature dependence of the conductance
around the stable weak tunneling fixed point.}
\end{figure}

In this case, the strong coupling fixed point is qualitatively different
from the previous cases. The minima are now arranged in a \emph{honeycomb
lattice} with minimal spacing $d=\frac{4\pi}{3}\frac{1}{\sqrt{2g}}$.
Therefore, the renormalization equation for the leading operator is

\begin{equation}
\frac{\partial\bar{\lambda}_{s}}{\partial\ell}=\left(1-\frac{2}{9g}\right)\bar{\lambda}_{s},\label{eq:dual_theory-1}
\end{equation}
and the dual theory is

\begin{equation}
\bar{H}_{s}=\bar{\lambda}_{s}\sum_{k=1}^{3}e^{i\sqrt{\frac{2}{9g}}\vec{a}_{k}.\vec{\Theta}\left(0\right)}s^{-}+h.c.\label{eq:dual_theory_H_S}
\end{equation}
where the $s^{\pm}$ operator connects the two sublattices of the
honeycomb lattice\citep{yi1996quantum}. In this regime, a \emph{stable
intermediate fixed point} emerges in the range: $\frac{2}{9}<g<\frac{1}{2}$. 

This fixed point can also be understood via a conformal embedding
to the boundary physics of the three state Potts model, where this intermediate
fixed point is associated to the $B+C$ boundary phase\citep{affleck2001quantum}.

While time-reversal symmetry in the transport problem suggests a connection
with the$\Gamma_{M}$ fixed point studied in \citet{oshikawa2006junctions},
the absence of Klein factor induced Hilbert space twisting here makes
it unlikely that the two fixed points are identical. Nevertheless,
for brevity and simplicity, I will denote this stable fixed point
at $2r\varphi\equiv\left\{ 0,\frac{2\pi}{3},\frac{4\pi}{3}\right\} \text{\,mod\,}2\pi$
also as $\Gamma_{M}.$

In the domain where Eq.\,(\ref{eq:dual_theory-1}) flows to strong
coupling, it is natural to consider the dual theory of Eq.\,(\ref{eq:dual_theory_H_S}).
The lattice of minima for the classical potential associated with
Eq.\,(\ref{eq:dual_theory_H_S}) is obtained by minimizing

\begin{equation}
\bar{F}=-\sqrt{3+2\sum_{k=1}^{3}\cos\left[\sqrt{\frac{2}{3g}}\vec{b}_{k}.\vec{\Theta}\left(0\right)\right]}.
\end{equation}

This recovers Eq.\,$\left(\ref{eq:Hs-boson}\right)$, and shows that
the theories become self-dual at $g=\frac{1}{3}$. In the quantum
Brownian motion analogy, this self-dual point corresponds to an identical
particle mobility in both descriptions, that is

\begin{equation}
\mu,\bar{\mu}=\frac{1}{2}.
\end{equation}

At $g=\frac{2}{9}$, one can go beyond the leading-order RG flow and
compute the next-order correction using an $\varepsilon$-expansion,
$\varepsilon=1-\frac{2}{9g}$. Due to inter-sublattice hopping, the
renormalization step generates the operators $\bar{\lambda}_{s}^{2}\vec{\alpha}_{k}.\partial_{x}\vec{\Theta}\left(0,0\right)s^{z}$.
This modifies the renormalization equation\citep{yi1996quantum} to

\begin{equation}
\frac{\partial\bar{\lambda}_{s}}{\partial\ell}=\varepsilon\bar{\lambda}_{s}-3\bar{\lambda}_{s}^{3}.
\end{equation}
The resulting stable fixed point is $\bar{\lambda}_{s}=\sqrt{\varepsilon/3}$
and the mobility at this point is \citep{yi1996quantum} 
\begin{equation}
\bar{\mu}=\pi^{2}\varepsilon.
\end{equation}

A similar analyze can be done around the $\Gamma_{0}$ fixed point
and close to $g=\frac{1}{2}$. In an $\epsilon$-expansion $\varepsilon=1-2g$,
the mobility is $\mu=1-\pi^{2}\varepsilon$. 

In the language of the Brownian motion, the picture that emerges from
this limiting cases is that the particle mobility increases steadily
from $1$ to 0 as $g$ changes from $\frac{2}{9}$ to $\frac{1}{2}$.
This behavior is depicted in Fig.\,\ref{fig:Renormalization-group-diagram.}. 

\section{Y-junction conductance \label{sec:Y-junction-conductance}}

There is an intrinsic contact resistance between the non-interacting
leads and any Luttinger liquid, which eliminates the usual $g$-dependence
found in conductance calculations using the Kubo formula. As a result,
a Luttinger liquid exhibits a measurable conductance of $\frac{e^{2}}{h}$,
regardless of its Luttinger parameter $g$. The same principle applies
to the conductance of any junction between Luttinger liquids: the
observable conductance is independent of the interaction strength.

This result is also physically intuitive. An electron or hole always
tunnels from the source lead into the Luttinger liquid and, after
traversing the circuit, must tunnel from the Luttinger liquid into
the drain lead. In other words, the charge and spin bosonic excitation
within the Luttinger liquid must recombine into a physical electron
or hole before entering the drain. While the internal transport processes
within the circuit can depend strongly on the Luttinger parameter,
the total measurable (two-terminal) conductance (defined from source
to drain) does not explicitly reflect this dependence, as it is constrained
by the properties of the non-interacting leads. As discussed in  \citet{oshikawa2006junctions},
the practical procedure is to: \emph{evaluate the conductance as a
function of $g$, find the correct fixed point that describes the
internal physics, and at the last step take $g=1$}.

It is convenient to decompose the conductance tensor into two distinct
components, following \citet{nayak1999resonant,oshikawa2006junctions}: 
\begin{itemize}
\item [${i}$:] $G_{s}$ correspondents to the symmetric part of the tensor,
it is the conductance when two terminals have no voltage applied to
them; 
\item [${ii}$:] $G_{A}$ is the anti-symmetric part of the tensor and is
zero for renormalization group fixed points with time reversal
symmetry. 
\end{itemize}
With these definitions, and accounting for\emph{ non-interacting leads},
the conductance tensor takes the general form:

\begin{equation}
G_{jk}^{s,c}=\left(3\delta_{jk}-1\right)\frac{G_{S}^{s,c}}{2}\pm\varepsilon_{jk}\frac{G_{A}^{s,c}}{2},\label{eq:conductance _tensor}
\end{equation}
with $\delta_{j,k}$ the Kronecker delta and $\varepsilon_{j,k}$
are the matrix elements of $\Omega-\Omega^{-1}$. 

The model of a $Y$-junction with edge states is time-reversal symmetric,
thus $G_{A}^{s,c}=0$. Therefore, characterizing the conductance tensor
reduces to determining the symmetric two-terminal conductance $G_{S}$.

The refermionized cases are very useful to understand the physics
of the $Y$-junction. In the non-interacting $g=1$, the unitarity
of the scattering matrix and the symmetry of the junction imposes
that the maximum conductance in a free fermion picture is \citep{nayak1999resonant}
\begin{equation}
G_{S}^{\text{MAX}}\leq\frac{8}{9}\frac{e^{2}}{h},
\end{equation}
per \emph{fermionic channel}. Hence, the maximum conductance for edge
states at $g=1$ in the absence of spin-flip events is 
\begin{equation}
G_{S}\left(g=1\right)\leq\frac{16}{9}\frac{e^{2}}{h}.
\end{equation}
The other two refermionize points correspond to correlated tunneling
events that separate charge and spin degrees of freedom. Hence, their
conductance are 

\begin{align}
G_{S}^{s}\left(g=\frac{1}{2}\right) & \leq\frac{8}{9}\frac{e^{2}}{h},
\end{align}
\begin{align}
G_{S}^{c}\left(g=2\right) & \leq\frac{8}{9}\frac{e^{2}}{h}.
\end{align}

For general values of $g$, the two terminal conductance, $G_{S}^{s,c}$,
can be evaluated in \emph{linear response} to an applied voltage on
one of the edge states\citep{nayak1999resonant},

\begin{widetext}
\begin{equation}
G_{S}^{s,c}=\lim_{\omega\to0^{+}}2\frac{e^{2}}{h}\left(1-\int_{-\infty}^{\infty}\int_{-L/2}^{L/2}d\tau dx\frac{e^{i\omega t}\left\langle J_{1}^{s,c}\left(y,\tau\right)J_{1}^{s,c}\left(x,0\right)\right\rangle }{2\pi\omega L}\right)
\end{equation}
\end{widetext}where $J_{j}^{s}=-\partial_{x}\theta_{j}$ is the\emph{
spin current}, and $J_{j}^{c}=\partial_{x}\phi_{j}$ is the \emph{charge
current} operator of the Luttinger liquid. 

At the $\Gamma_{0}$ fixed point, 

\begin{equation}
\int_{-\infty}^{\infty}\int_{-L/2}^{L/2}d\tau dx\frac{e^{i\omega t}\left\langle J_{1}^{s,c}\left(y,\tau\right)J_{1}^{s,c}\left(x,0\right)\right\rangle }{2\pi\omega L}=1
\end{equation}
and the conductance vanishes,
\begin{equation}
G_{S}^{s,c}\left(\Gamma_{0}\right)=0
\end{equation}
 To evaluate the correlation function at $\Gamma_{A}$ fixed point,
the current operators must first be expressed in terms of the rotated
fields, Tab.\,\ref{tab:boson_def}. When $\Phi_{1,2}/\Theta_{1,2}$
are \emph{pinned}, their fluctuations are suppressed, and the corresponding
expectation values vanish. Thus, the conductance at $\Gamma_{A}$
is

\begin{equation}
G_{S}^{s,c}\left(\Gamma_{A}\right)=\frac{4}{3}\frac{e^{2}}{h},
\end{equation}
that exceeds the unit of conductance due to an Andreev reflection\citep{nayak1999resonant}.

The final stable renormalization group fixed point is $\Gamma_{M}$
(defined on sub-Sec.\,\ref{subsec:gamma_M}). As an intermediate
fixed point, evaluating the current-current correlation function is
generally nontrivial. However, analytical results are available for
special values of $g$.

Near to $g=\frac{1}{2}$ and $\frac{2}{9}$ the $\varepsilon$ expansions
gives a conductance that is a $\varepsilon$ deviation from $G_{S}^{s,c}\left(\Gamma_{0}\right)$
and $G_{S}^{s,c}\left(\Gamma_{A}\right)$ respectively. At the self
dual point, $g=\frac{1}{3}$, the mobility $\mu=\frac{1}{2}$ implies 

\begin{equation}
\int_{-\infty}^{\infty}\int_{-L/2}^{L/2}d\tau dx\frac{e^{i\omega t}\left\langle J_{1}^{s}\left(y,\tau\right)J_{1}^{s}\left(x,0\right)\right\rangle }{2\pi\omega L}=\frac{2}{3}.
\end{equation}
Hence the conductance is

\begin{equation}
G_{S}^{s}\left(\Gamma_{M},g=\frac{1}{3}\right)=\frac{2}{3}\frac{e^{2}}{h}.
\end{equation}

The picture that emerges is that the conductance for $\Gamma_{M}$
evolves \emph{smoothly} from the $G_{S}^{s}\left(\Gamma_{A}\right)$
value at $g=\frac{2}{9}$ to zero at $g=\frac{1}{2}$.

\section{Discussion and Conclusions\label{sec:Discussion-and-Conclusions}}

This work explored the transport properties of a Y-junction formed by three
interacting helical edge states in a two-dimensional topological insulator (2DTI). 
Using bosonization and duality mappings, I uncovered a rich phase diagram governed
by the interplay between tunneling phases, $r\phi$, and interaction strength, $g$.
The system exhibits multiple renormalization group (RG) fixed points
that control its low-energy behavior, including a novel intermediate
fixed point arising for strong repulsive interactions.
These findings demonstrate how correlated tunneling processes
can qualitatively modify transport in multiterminal topological devices.

The single-particle tunneling phase, $r\phi$, is inherited from the microscopic Kane and Mele model.
While the spin-orbit interaction of the material define $\phi$,
the parameter  $r=\frac{d_0}{\sqrt{3}a}$ depends on the junction geometry, with $d_0$ 
denoting the distance between edges and $a$ the lattice constant.
Both parameters play a crucial role in determining the conductance of the Y-junction.

For example, consider the case $\phi=\pi/2$ and the geometry shown of Fig.~\ref{fig:Microscopic-model-Y},
where the edge separation is $d_0 = 2\sqrt{3}a$.
For this configuration $d_0=2\sqrt{3}a$, 
and the intermediate
fixed point for $g<1/2$ is stable (Fig.~\ref{fig:Renormalization-group-diagram.}).
Conversely, if the junction geometry yields $d_0=\sqrt{3}a$, 
then $2r\phi \not\equiv 0 \mod 2\pi$, and the intermediate fixed point becomes unstable (Fig.~\ref{fig:Renormalization-group-diagram}).

The renormalization
group flows are summarizes in the Figs.\,(\ref{fig:Renormalization-group-diagram} and \ref{fig:Renormalization-group-diagram.}),
showing that the system is governed by two unstable and three stable
fixed points ($\Gamma_{0}$, $\Gamma_{A}$ and $\Gamma_{M}$). 

In the case of weak interactions, $\frac{2}{3}<g<\frac{3}{2}$, the
conductance of the junction is always zero. In this range, no tunneling
process is relevant at any coupling constant. However, at finite temperature,
there are corrections to the conductance proportional to the temperature
to the power of the leading irrelevant operator, Eqs.\,(\ref{eq:finte-temp-1})-(\ref{eq:finte-temp-3}). 

For intermediate and strong interactions, the behavior of the system
depends on the tunneling phase:
\begin{itemize}
\item [$i$:] Generic phase: when $2r\varphi\not\equiv\left\{ 0,\frac{2\pi}{3},\frac{4\pi}{3}\right\} \text{\,mod\,}2\pi$
and at intermediate values of the interaction, $\frac{1}{2}<g<\frac{2}{3}$
or $\frac{3}{2}<g<2$, there is a critical coupling constant, $\lambda_{s,c}^{\text{critical}}\propto\left(\lambda_{e}^{\text{critical}}\right)^{2}$.
For bare coupling values above this threshold, the system will flow
towards the $\Gamma_{A}$ fixed point. This corresponds to the ``Andreev
reflection'' fixed point and the spin or charge conductance is $G_{S}^{s,c}=\frac{4}{3}\frac{e^{2}}{h}$.
In the strongly interacting regime, $g<\frac{1}{2}$ or $g>2$, no
critical coupling is required, and the system always flows to $\Gamma_{A}$.
\item [$ii$:] For $2r\varphi\equiv\left\{ 0,\frac{2\pi}{3},\frac{4\pi}{3}\right\} \text{\,mod\,}2\pi$
and strongly repulsive interactions, $\frac{2}{9}<g<\frac{1}{2}$,
the system flows to an intermediate fixed point, $\Gamma_{M}$. Where

the spin conductance evolves \emph{smoothly} from the $G_{S}^{s}\left(\Gamma_{A}\right)$
value at $g=\frac{2}{9}$ to zero at $g=\frac{1}{2}$.
\end{itemize}

For simplicity, I consider only symmetric tunneling amplitudes.  
In the case of asymmetric couplings, the strong-coupling ($\Gamma_A$) and weak-coupling ($\Gamma_0$)
fixed points naturally restore the $Z_3$ symmetry.  
The existence of an intermediate stable fixed point follows the same reasoning as before,
although its conductance may differ from the $\Gamma_M$ with symmetric couplings.  
Therefore, as shown in Ref.~\citet{oshikawa2006junctions}, the phase diagram of the
Y-junction remains qualitatively unchanged in the presence of asymmetries.

The time-reversal symmetry inherent to the Y-junction geometry considered
in Fig.\,\ref{fig:Microscopic-model-Y} precludes the two chiral fixed
point $\chi_{\pm}$ described in \citet{chamon2003junctions,oshikawa2006junctions}.
These fixed points are of particular interest for applications as
spin filters or charge transistors. They do exist as  particular
values of the non-interacting, $g=1$, and refermionized cases $g=\left\{ \frac{1}{2},2 \right\}$,
but it is still an open question whether alternative geometries could
be used to generate and stabilize them in the context of a general $g$.

In summary, this work has explored the transport properties and phase
diagram of a Y-junction formed by three interacting edge states in
a topological insulator. By analyzing the renormalization group flows
and evaluating the conductance tensor, the study maps out the rich
interplay between tunneling phases and interaction strength, and identifies
distinct fixed points that govern the low-energy behavior of the system.
\begin{acknowledgments}
This work was partially supported by S\~{a}o Paulo Research Foundation (FAPESP), Brazil, Process No. 2022/15453-0.
The author would like to thank I. Affleck, whose work underpins many of
the results discussed in this manuscript. The author also would like to thank H.
Blanc for insightful discussions and consistent encouragement.
\end{acknowledgments}

\end{document}